\documentclass[%
 reprint,
superscriptaddress,
 amsmath,amssymb,
 aps, physrev,
]{revtex4-2}

\usepackage{siunitx}
\usepackage{cases}
\usepackage{soul}

\usepackage{graphicx}
\usepackage{dcolumn}
\usepackage{bm}
\usepackage{hyperref}
\usepackage{comment}
\hypersetup{
    colorlinks=true,
    linkcolor=blue,
    citecolor=blue,
    filecolor=blue,
    urlcolor=blue,
}

\newcommand*{\rom}[1]{\expandafter\@slowromancap\romannumeral #1@}
\def\BigRoman{\uppercase\expandafter{\romannumeral\number\count 255 }}
\def\Romannumeral{\afterassignment\BigRoman\count255=}
\def\Supplements{Supplemental Material}
\usepackage[mathlines]{lineno}

\begin{document}

\title{Molecular Origin of UV-Induced Irreversible Phase Changes in a Chromonic Liquid Crystal}

\author{Junghoon Lee}
\affiliation{
 Department of Physics, Ulsan National Institute of Science and Technology, Ulsan, Republic of Korea
}
\author{Seonghun Jeong}
\affiliation{
 Department of Chemistry, Ulsan National Institute of Science and Technology, Ulsan, Republic of Korea
}
\author{Jung-Min Kee}
\email{jmkee@unist.ac.kr}
\affiliation{
 Department of Chemistry, Ulsan National Institute of Science and Technology, Ulsan, Republic of Korea
}
\author{Joonwoo Jeong}
\email{jjeong@unist.ac.kr}
\affiliation{
 Department of Physics, Ulsan National Institute of Science and Technology, Ulsan, Republic of Korea
}
\affiliation{
 UNIST Research Center For Soft and Living Matter, Ulsan National Institute of Science and Technology, Ulsan, Republic of Korea
}

\date{\today}

\begin{abstract}
Aqueous solutions of disodium cromoglycate (DSCG), a representative model system for chromonic liquid crystals, exhibit temperature- and concentration-dependent phase behaviors spanning isotropic, nematic, and columnar phases, as well as their coexistence regions. Nastishin et al. (2018) \cite{Nastishin2018} reported that UV irradiation can alter the phase diagram, transforming a nematic phase into a nematic-isotropic biphasic state due to weakened molecular attractions, accompanied by a slow post-irradiation relaxation. Here, we revisit this phenomenon and elucidate the molecular origin of this phase diagram shift: the UV-induced photodegradation of DSCG into specific photodegradation products, which we identify using liquid chromatography-mass spectrometry. Through an integrated approach combining \textit{in situ} X-ray scattering and polarized optical microscopy, we demonstrate that these degradation products disrupt the self-assembly of DSCG aggregates, thereby expanding the isotropic and biphasic regions in the phase diagram. These findings demonstrate that chromonic assemblies and their phase behaviors are highly sensitive to minor chemical alterations, providing a potential route toward light-controlled self assembly of soft matter.
\end{abstract}

\maketitle


\section{Introduction}\label{sec:intro}

Lyotropic chromonic liquid crystals (LCLCs) comprise a class of aqueous self-assembled systems in which aromatic molecules form polydisperse $\pi$-stacked aggregates that give rise to liquid crystal mesophases.\cite{TamChang2008, Lydon2010} Unlike surfactant-based assemblies \cite{Tiddy1980}, chromonics lack a critical micelle concentration and instead exhibit continuously evolving aggregate lengths governed primarily by concentration and temperature. Their phase behavior is further modulated by additives that alter electrostatic screening or hydration environments,\cite{Kostko2005,Zhang2016,Asdonk2017,Ibanez2018,Koizumi2019,Lee2019,eunLyotropicChromonicLiquid2020,Cheon2025} by mixed chromonic species with distinct stacking propensities,\cite{Yamaguchi2016,Rivas2017} and by co-solvents that influence hydrophobic interactions.\cite{Hahn1998}

Disodium cromoglycate (DSCG) is a widely used model compound for probing the fundamental aggregation and mesophase formation of LCLCs.\cite{Goldfarb1985,Nastishin2005} Owing to its chromone moiety, DSCG exhibits strong UV absorption,\cite{Vasyuta2005,Yusuf2019} and its optical anisotropy is known to respond to irradiation.\cite{Nastishin2018} Energetic considerations have suggested that UV photon energies exceed the estimated scission energy of $\pi$-stacked aggregates, motivating the view that irradiation primarily shortens aggregates without altering molecular structure.\cite{Hunter1990,Ringer2006} Within this framework, UV-induced changes are expected to relax on thermal-equilibration timescales characteristic of isodesmic aggregation.

However, the UV response of DSCG has not been systematically verified against these assumptions. Importantly, a purely physical scission mechanism would yield reversible variations in birefringence, whereas our observations show that the UV-induced changes persists far beyond the thermal relaxation timescale. This discrepancy indicates that additional irreversible processes contribute to the optical and structural evolution of the irradiated samples.

In this study, we examine the effect of UV irradiation on the phase behavior of aqueous DSCG solutions. Liquid chromatography and mass spectrometry reveal the formation of photodegradation products that act as impurities \cite{Mansfield1999,Ali2008,El-Bagary2016,Zhu2017a}, thereby modifying nematic ordering and aggregate structures. Polarized optical microscopy and X-ray scattering further demonstrate that the UV irradiation leads to persistent changes in the phase diagram. These results establish that the UV response of DSCG is dominated not by reversible aggregate scission, but by chemical alterations to the DSCG molecules, offering a framework for understanding photochemical influences on chromonic self-assembly.

\section{Results and Discussion}\label{sec:rnd}

\begin{figure}[t!]
\centering
\includegraphics{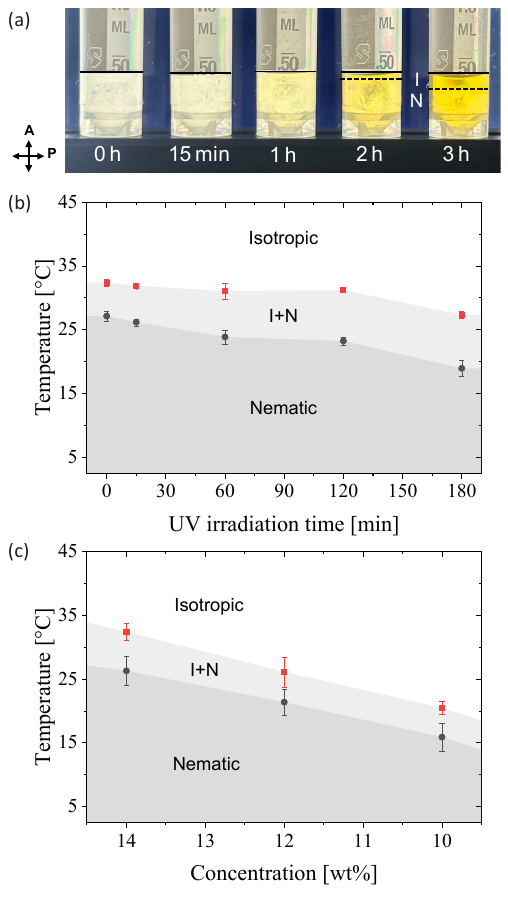}  
\caption{\label{fig:phase} 
UV-induced changes in phase behavior. 
(a) Photographs of UV-irradiated DSCG solution contained in tubes according to the exposure time. The tubes are placed between two crossed polarizers. The color turns yellowish, and after 2 hours, an interface shown as a dashed line appears between the nematic phase (N) at the bottom and the isotropic phase at the top. The isotropic region expands with more UV exposure.
(b) Phase-transition temperatures according to UV irradiation time. The biphase(B) regime (I+N) widens and shifts downward. Black circles represent the phase transition temperature $T_{\text{NB}}$ and red squares represent $T_{\text{BI}}$.
(c) The phase diagram of the neat DSCG solution. As the concentration decreases, the biphase region shifts downward but narrows. The horizontal axis is reversed, so that concentration increases from right to left, for direct comparison with (b).}
\end{figure}

UV irradiation alters the phase behavior of DSCG. When exposed to UV light over time, nematic DSCG at room temperature enters the nematic-isotropic biphase without a temperature change.
Fig.~\ref{fig:phase}(a) shows that the isotropic region grows as the irradiation time increases, reaching approximately one-third of the initial sample after 3 hours of UV exposure ($\lambda=365~\si{\nano\meter}$; 164.42~\si{mW/cm^2}). While this UV-induced melting was reported in \cite{Nastishin2018}, changes in the phase transition temperature have not been systematically studied.
Fig.~\ref{fig:phase}(b) shows UV-induced changes in two phase-transition temperatures: $T_{\text{NB}}$, from the nematic phase (N) to a biphase (B) of isotropic and nematic phase, and $T_{\text{BI}}$, from the biphase to the isotropic phase (I). \ref{sec:mnm} describes detailed procedures for determining the transition temperatures. Both $T_{\text{NB}}$ and $T_{\text{BI}}$ decrease monotonically as UV irradiation time increases.

UV-induced changes in phase behavior contrast with DSCG concentration-induced changes and hint at a compositional change. 
The phase diagram of UV-exposed 14 wt\% DSCG shown in Fig.~\ref{fig:phase}(b) does not overlap with the phase diagram of neat DSCG solution shown in Fig.~\ref{fig:phase}(c). Specifically, the biphasic range widens with increasing UV exposure time, whereas a decrease in the DSCG concentration in the absence of UV exposure narrows the biphasic range. 
This finding suggests that UV irradiation may affect the composition of aqueous DSCG solution, e.g., via a photochemical reaction. The UV-induced changes in appearance---UV-exposed specimen becomes yellowish as shown in Fig.~\ref{fig:phase}(a)---and the UV-Vis spectrum in the \Supplements~support this hypothesis.

\begin{figure}[t]
\centering
  \includegraphics{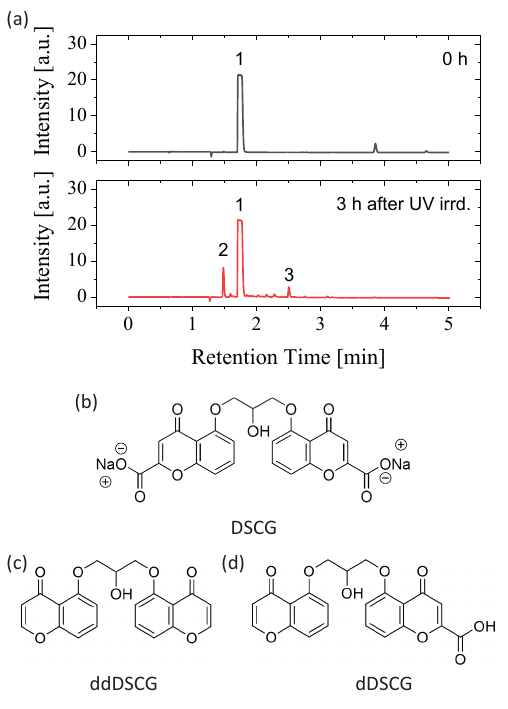}  
  \caption{\label{fig:photochem} 
(a) LC-MS chromatograms of DSCG solution. The chromatogram at the top is from the neat DSCG, and the bottom one from the DSCG after 3-hr UV irradiation. Peak no. 1 corresponds to the DSCG; peaks nos. 2 and 3 appear after UV exposure. 
(b) Molecular structure of DSCG. 
(c and d) Molecular structure of UV irradiation-induced photodegradation products corresponding to peaks no. 2 and 3 in (a). Two carboxyl groups of (a) are missing in (c), and one is missing in (d).}
\end{figure}

\begin{figure*}[t!]
\centering
\includegraphics{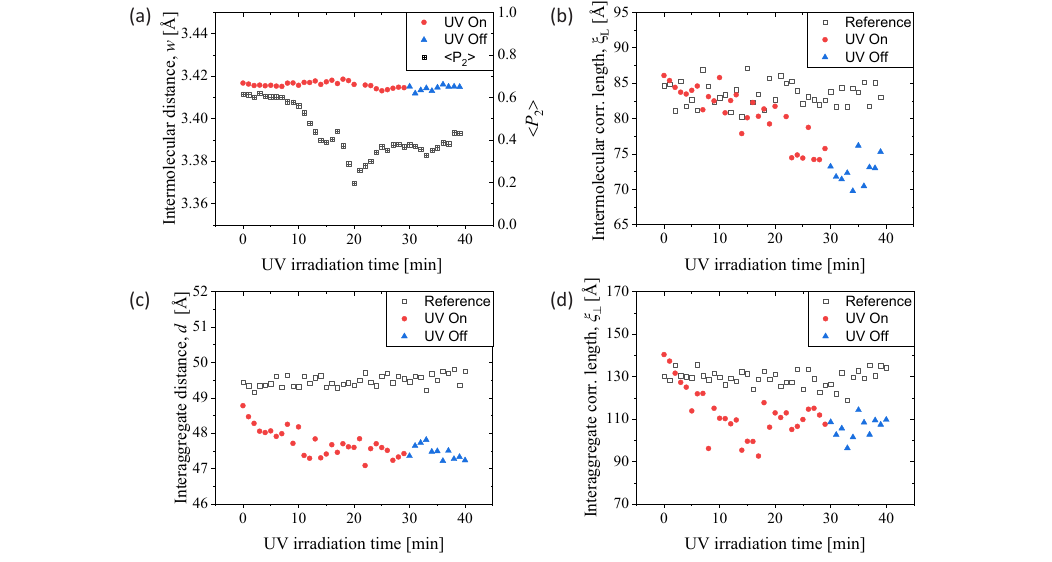}  
\caption{\label{fig:xray} 
Microstructural changes upon UV irradiation.
(a) Changes in intermolecular distance $w$ and order parameter $<P_2>$ according to UV irradiation time. Red circles indicate $w$ when UV light is on, and blue triangles indicate $w$ after turning off the UV light. Hollow squares represent orientational order parameter $<P_2>$ calculated from the same WAXD data.
Changes in (b) the intermolecular correlation length $\xi_{\text{L}}$, (c) interaggregate distance $d$, and (d) the interaggregate correlation length $\xi_\perp$ according to UV irradiation time. Black squares show reference data obtained from a UV-free specimen measured under X-ray exposure to assess beam-induced changes.
}
\end{figure*}

Liquid chromatography-mass spectroscopy (LC-MS) of UV-exposed specimens confirms this UV-induced change in composition. Comparing chromatograms of neat and UV-exposed DSCG, we first find a UV-induced decrease in the area under the curve of the peak corresponding to the DSCG (See the \Supplements). Additionally, UV irradiation results in two new peaks in the chromatogram. Mass spectra identify the two major photodegradation products shown in Fig.~\ref{fig:photochem}(c) and (d): the DSCG-like molecules with one (decarboxyDSCG, dDSCG) or two carboxyl groups (didecarboxyDSCG, ddDSCG) removed from DSCG. In short, UV irradiation cleaves the carboxyl groups of some DSCG molecules, lowering the effective DSCG concentration and producing two distinct types of photodegradation products.

In-situ SAXS/WAXD measurements reveal how UV exposure affects the microstructure of DSCG mesophase. While the intermolecular distance $w \approx 3.4~\si{\angstrom}$ is unaffected by the UV irradiation, as shown in Fig.~\ref{fig:xray}(a), the order parameter $<P_2>$ decreases and does not recover after turning off the UV light. This weakening of nematic ordering may result from a reduction in mean aggregation length $<\text{L}>$, often interpreted from the intermolecular correlation length $\xi_{\text{L}}$ \cite{Collings2010}. Indeed, Fig.~\ref{fig:xray}(b) presents that $\xi_{\text{L}}$ also begins to decrease when UV light is turned on and does not recover after UV light is turned off.

The SAXS results are consistent with the UV-induced shortening of the aggregates. As shown in Fig.~\ref{fig:xray}(c) and (d), the interaggregate distance $d$ and its correlation length $\xi_\perp$ also decrease with UV irradiation. Because the total number of molecules, including neat and damaged DSCG, is conserved, shorter but more number of aggregates lead to a shorter interaggregate distance $d$. In contrast, if the total DSCG concentration simply decreases, $\xi_{\text{L}}$ would decrease, but $d$ would not. This result indicates that DSCG-like photodegradation products identified by LC-MS participate in aggregate formation but affect the microstructure, e.g., limiting aggregate growth due to weak intermolecular interactions. 

\begin{figure*}[t!]
\centering
\includegraphics[width=\textwidth]{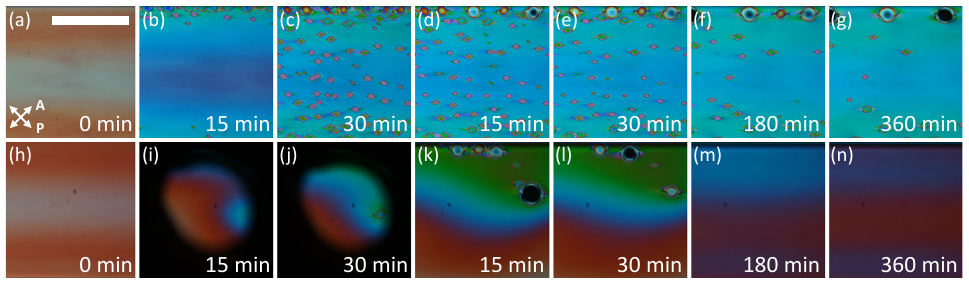}  
\caption{\label{fig:optics} 
Optical changes induced by global and local UV irradiation. 
The sample on the top row is exposed to UV light for 30 min, over the entire sample area (global irradiation), while the bottom one through an aperture (local irradiation). 
(a–c) and (h–k) show the evolution of birefringence and the growth of isotropic tactoids during UV exposure. 
The dark regions in (i) and (j) correspond to an aluminum mask with a small opening; oblique UV illumination can affect regions beneath the mask.
(d–g) and (k–n) show the post-irradiation evolution. 
All images are acquired with crossed polarizers of which pass axes are shown with white double arrows in (a). 
The white scale bar indicates 50~\si{\micro\meter}.
}
\end{figure*}

Lastly, comparison of optical texture changes by global and local UV irradiation supports our hypothesis that the photodegradation products alter the phase diagram. 
Fig.~\ref{fig:optics}(a–g) shows the evolution of birefringence and the growth of isotropic tactoids upon UV exposure over the entire sample area, i.e., the global irradiation. 
Although the number of isotropic tactoids decreases over time after the UV illumination is turned off, the tactoids persist, which is consistent with the change in $T_{\text{NB}}$ from $27.41\pm0.36\si{\degreeCelsius}$ to $20.62\pm0.91\si{\degreeCelsius}$. 
This provide clear evidence of a permanent change in sample composition and dispute the scenario that chromonic aggregates exhibit UV-induced transient changes and post-irradiation recovery \cite{Nastishin2018}.
On the other hand, Fig.~\ref{fig:optics}(h–n) shows the evolution after the local UV irradiation. 
UV-generated photodegradation products in the irradiated region diffuse out, and the sample recovers the fully nematic texture; the change in the interference color hints at the change in birefringence, i.e., sample composition.

\section{Materials and methods}\label{sec:mnm}
\subsection{Sample preparation and UV irradiation}
Disodium cromoglycate (DSCG) was purchased from Sigma-Aldrich and used as received (99.7\% purity). 
DSCG solutions were prepared by dissolving the DSCG powder in deionized water (18.2~\si{\mega\ohm\cm}). 
For UV-irradiated samples used in HPLC and LC-MS, we placed a 14.0 wt\% DSCG solution in a quartz cuvette (path length = 10 mm) under a 365-nm LED lamp (Liimtech, Republic of Korea) equipped with a diffuser. 
Before aliquoting samples from the cuvette, the solution was homogenized by repeated pipetting. Note that UV-irradiated biphase samples underwent macroscopic phase separation by density difference. To clearly visualize the isotropic–nematic interface, the sample tubes were briefly centrifuged prior to imaging.
We controlled the UV intensity to 164~\si{\milli\watt\per\square\centi\meter}. 
During UV illumination, the cuvette was sealed with a Teflon cap to prevent water evaporation.

\subsection{Determination of phase transition temperature}
We determined the phase transition temperatures of the samples by examining their optical textures under polarized light microscopy (POM).
UV-irradiated DSCG solutions (14.0 wt\%, collected from a quartz cuvette) were loaded into cylindrical capillaries (CV1017, Vitrocom) with a diameter of 100~\si{\micro\meter} ($\pm 10\%$ tolerance) and sealed with epoxy. Prior to loading, the solutions were homogenized by repeated pipetting. 
Four characteristic transition temperatures were identified based on optical texture changes: $T_{\text{NB}}$, $T_{\text{BI}}$, $T_{\text{IB}}$, and $T_{\text{BN}}$. 
The nematic (N) to biphase (B) transition temperature $T_{\text{NB}}$ is the temperature at which an isotropic tactoid first appears upon heating at 0.5\si[per-mode = symbol]{\degreeCelsius\per\minute}. 
The biphase (B) to isotropic (I) transition temperature $T_{\text{BI}}$ is the temperature at which the sample starts to appear completely dark in POM with crossed polarizers. 
The temperature $T_{\text{BN}}$, from biphase to nematic, and $T_{\text{IB}}$, from isotropic to biphase, are determined similarly;  all isotropic tactoids disappear at $T_{\text{BN}}$, and bright textures begin to appear at $T_{\text{IB}}$ upon cooling. 
All measurements were conducted on a temperature-controlled stage (LTS120/T96, Linkam Scientific Instruments, UK). 
Before observation, samples were stored at 65\si{\degreeCelsius} for 1 hour, cooled to 22\si{\degreeCelsius} at 20\si[per-mode = symbol]{\degreeCelsius\per\minute}, and equilibrated for 4 hours.

\subsection{HPLC analysis for the quantification of DSCG}
Samples were aliquoted from the cuvette and homogenized by repeated pipetting prior to analysis. Samples were analyzed by reverse-phase HPLC (Agilent 1260 Infinity \MakeUppercase{\romannumeral 2}) using Eluent A (water + 10 mM NH$_4$OAc) and Eluent B (CH$_3$CN:water = 9:1 + 10 mM NH$_4$OAc). A gradient of 0--70\% B over 25 min was applied at a flow rate of 1 mL/min with UV detection at 214, 254, and 280 nm. A C18 Eclipse column (4.6$\times$150 mm) was used at 30\si{\degreeCelsius}.

\subsection{LC-MS analysis for the identification of photodegradation product}
Samples are analyzed by LC-MS (Waters Xevo G2 QTof). The eluents are the same as those used in the HPLC experiment. The gradient was programmed as follows: 0--5 min, 100\% A to 30\% A; 5--7 min, 30\% A; 7.1--10 min, 100\% A. It was applied at a flow rate of 0.45 mL/min with UV detector. ACQUITY UPLC HSS C18 column(1.8 \si{\micro\meter}, 2.1$\times$100 mm, Waters) was used at 30\si{\degreeCelsius}.

\subsection{Microstructure analysis from in-situ SAXS/WAXD}
We performed in-situ small-angle X-ray scattering (SAXS) and wide-angle X-ray diffraction (WAXD) experiments at the PLS-\Romannumeral 2 6D UNIST-PAL beamline of the Pohang Accelerator Laboratory, Republic of Korea. The X-ray energy was 11.564 keV, covering a $q$-range of 0.0065 to 3.35~\si{\per\angstrom}. 
We adjusted the X-ray exposure time to 10-20 seconds for single-shot acquisition to acquire enough signal-to-noise ratio while minimizing the X-ray damage to specimens. As shown in Fig.~\ref{fig:xray}, the structural change by X-ray, shown in the Reference data without UV, is minor compared with the structural changes induced by UV irradiation.
We prepared homogeneously aligned nematic DSCG samples confined between two parallel rubbed borosilicate coverslips with a spacing of 120~\si{\micro\meter}. 
Epoxy sealing minimized the water evaporation.
We verified homogeneous alignment optically prior to the X-ray measurements. 
The sample was mounted perpendicular to the incident X-ray beam, while polarized and collimated 365-\si{\nano\meter} UV LED light (M365L3, Thorlabs) was directed onto the X-ray probed region of sample at 16\si{\degree} from 15~\si{\centi \meter} during measurements. The UV polarization axis was perpendicular to the nematic alignment direction.
The UV intensity at the sample location was 0.55~\si{\milli\watt\per\square\centi\meter}. Note that the UV intensity used for phase-diagram mapping (Fig.~\ref{fig:phase}) and chemical analysis was much higher, 164~\si{\milli\watt\per\square\centi\meter}, reflecting different experimental constraints, e.g., optical path length.
All X-ray data were collected at 22\si{\degreeCelsius}.

The average inter-aggregate spacing $d$ and inter-aggregate correlation length $\xi_\perp$, as well as the inter-molecular spacing $w$ and inter-molecular correlation length $\xi_{\text{L}}$, were estimated from the SAXS and WAXD peaks, respectively \cite{eunLyotropicChromonicLiquid2020,Cheon2025}. 
The position and full width at half maximum (FWHM) of the small-angle peak near $q=0.12$~\si{\per\angstrom} yielded $d$ and $\xi_\perp$, while the wide-angle peak near $q=1.83$~\si{\per\angstrom} yielded $w$ and $\xi_L$. 
The orientational order parameter $\langle P_2\rangle=\langle (3\cos^2\beta-1)/2\rangle$ were computed from the Kratky kernel $\beta$ of azimuthal intensity distribution $I(\theta)$ \cite{Davidson1995,Agra-Kooijman2018}.

\subsection{Microscopic observation under global and local UV irradiation}

We performed \textit{in situ} optical experiments using polarized optical microscopy (POM). DSCG samples (14.0 wt\%) were loaded into rectangular borosilicate capillaries (5010, Vitrocom) at room temperature and sealed with epoxy. The nominal path length of capillaries was 100~\si{\micro\meter} ($\pm 10\%$ tolerance). Before UV irradiation in the same capillaries, the samples were allowed to relax at room temperature for at least 4 hours to obtain a homogeneous optical texture in the ground state. UV light (0.55~\si{\milli\watt\per\square\centi\meter}) was applied simultaneously with image acquisition during POM observation. To keep the imaging light path clear, the UV LED light (M365L3, Thorlabs) was directed obliquely at an angle of 35\si{\degree} and positioned 15~\si{\centi\meter} from the sample. The UV light was collimated, its polarization axis was perpendicular to the nematic alignment direction. For local irradiation, an aluminum foil mask perforated with a 25-gauge hole (5/8'' diameter) was used. All optical data were acquired at 22.0\si{\degreeCelsius} on a temperature-controlled stage (LTS120/T96, Linkam Scientific Instruments, UK).

\section{Conclusion}\label{sec:conc}
We examine the effects of UV irradiation on the phase behavior of DSCG. By combining optical microscopy, synchrotron X-ray scattering, and LC-MS, we characterize the phase behavior and microstructure of UV-irradiated DSCG and find that photochemical byproducts are responsible for the irreversible changes. The photodegradation products missing one or two carboxyl groups seem to act as impurities, leading to lower phase-transition temperatures and shorter aggregates with decreased correlation lengths. Looking forward, isolating or synthesizing the photodegradation products, followed by systematic doping experiment, elucidate their microscopic roles in phase behaviors.

Our findings not only highlight the sensitivity of chromonic mesophases to minor chemical alterations \cite{eunLyotropicChromonicLiquid2020} but also propose a potential route for the optical control of soft matter self-assembly. Photodegradation products, including potentially chiral species, or photoresponsive additives may serve as \textit{in situ} regulators of aggregation and phase behaviors, enabling light-tunable self-assembly.

\begin{acknowledgments}
The authors gratefully acknowledge financial support from the National Research Foundation of Korea: RS-2024-00345749 (J.L. and J.J.) and RS-2022-NR069957 (J.-M.K.). The X-ray scattering experiments were performed at the PLS-\MakeUppercase{\romannumeral 2} 6D UNIST-PAL Beamline of Pohang Accelerator Laboratory (PAL) in the Republic of Korea (proposal number 2023-3rd-6D-A013). We thank UNIST Office of Research Facilities and Training (ResFacT) for their equipment support with LC-MS. Junsoo Jang is acknowledged for the initial exploratory work.
\end{acknowledgments}

\end{document}



\title{Supplemental Material for ``Molecular Origin of UV-Induced Irreversible Phase Changes in a Chromonic Liquid Crystal''}

\author{Junghoon Lee}
\affiliation{
 Department of Physics, Ulsan National Institute of Science and Technology, Ulsan, Republic of Korea
}
\author{Seonghun Jeong}
\affiliation{
 Department of Chemistry, Ulsan National Institute of Science and Technology, Ulsan, Republic of Korea
}
\author{Jung-min Kee}
\email{jmkee@unist.ac.kr}
\affiliation{
 Department of Chemistry, Ulsan National Institute of Science and Technology, Ulsan, Republic of Korea
}
\author{Joonwoo Jeong}
\email{jjeong@unist.ac.kr}
\affiliation{
 Department of Physics, Ulsan National Institute of Science and Technology, Ulsan, Republic of Korea
}
\affiliation{
 UNIST Research Center For Soft and Living Matter, Ulsan National Institute of Science and Technology, Ulsan, Republic of Korea
}

\date{\today}

\maketitle

\section{Absorption spectra of samples}

\begin{figure}[h]
\centering
  \includegraphics{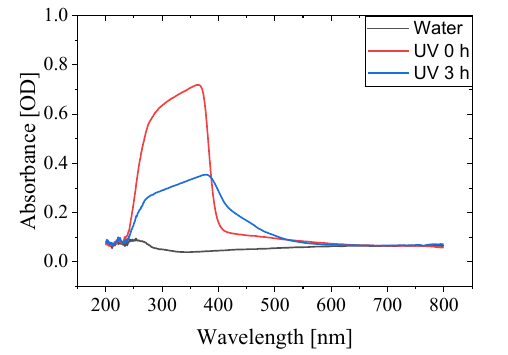}  
  \caption{\label{fig:uvvis} 
Changes in UV-Vis absorption spectra. Red and blue lines show measurements from neat and UV-irradiated (3 hr) sample cells, respectively. The black line shows deionized water in the same container as a reference.}
\end{figure}

\section{Area under the curve of HPLC and mass spectra of LC-MS}

\begin{figure}[h]
\centering
  \includegraphics{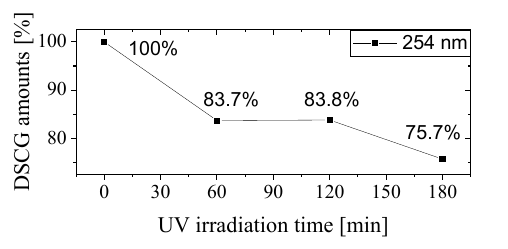}  
  \caption{\label{fig:hplc} 
Area under the curve (AUC) of the DSCG peak in the HPLC as a function of UV irradiation time. AUC, reflecting the amounts of DSCG, decreases as the UV irradiation time increases.}
\end{figure}

\begin{figure}[h]
\centering
  \includegraphics{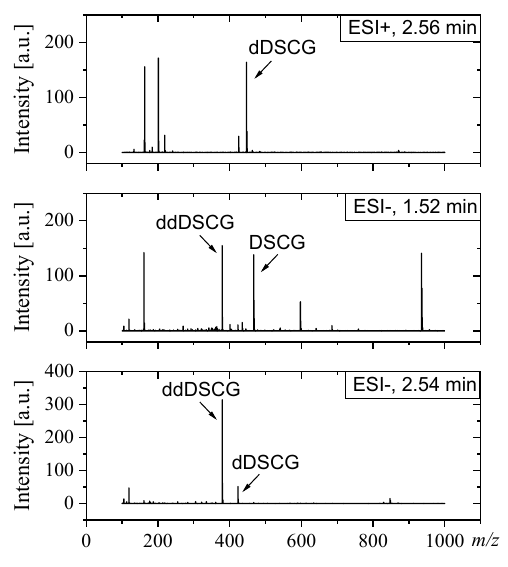}  
  \caption{\label{fig:mass} 
LC-MS data of the DSCG sample after 10 hours of UV irradiation. We identify two photodegraded compounds: ddDSCG and dDSCG.}
\end{figure}

\section{X-ray scattering data}

\begin{figure}[h]
\centering
  \includegraphics{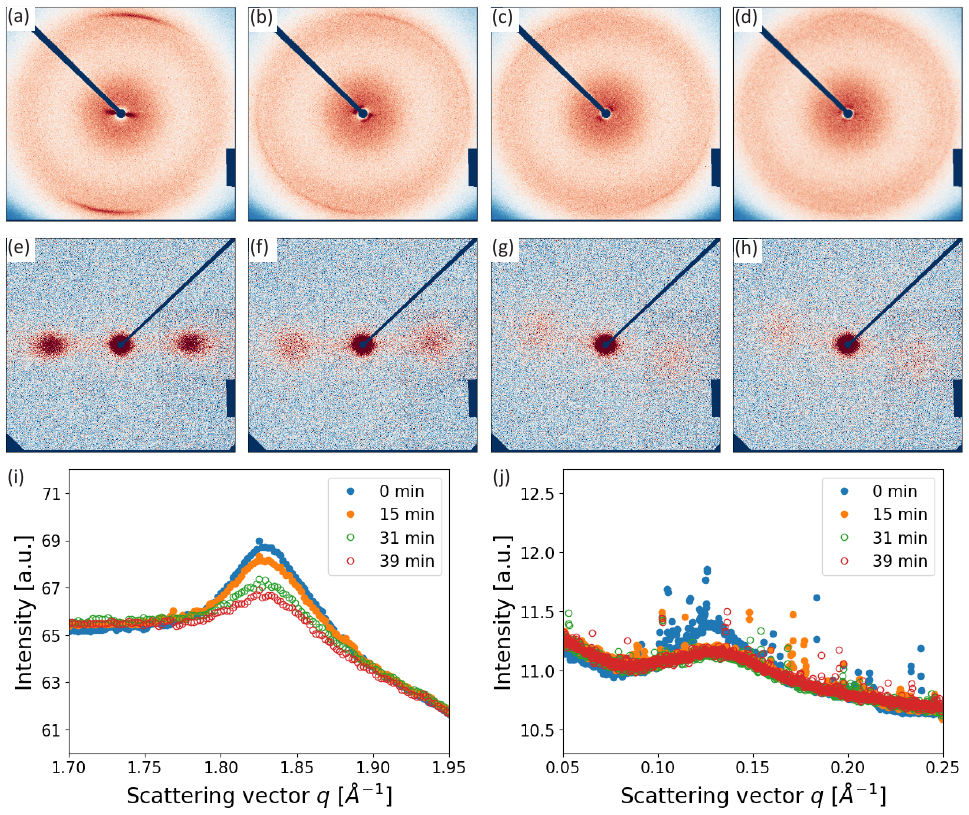}  
  \caption{\label{fig:raw-global} 
Comparison of X-ray scattering data for samples subjected to UV irradiation. The entire sample area was uniformly illuminated. Panels (a,e) correspond to 0 min, (b,f) to 15 min, (c,g) to 30 min, and (d,h) to 39 min of irradiation. (a–d) In the wide-angle region, the scattering peak broadens while the intensity decreases compared to the neat sample shown in (a). (e–h) A similar trend is observed in the small-angle region. (i,j) show the azimuthally averaged intensity profiles.}
\end{figure}

\begin{figure}[h]
\centering
  \includegraphics{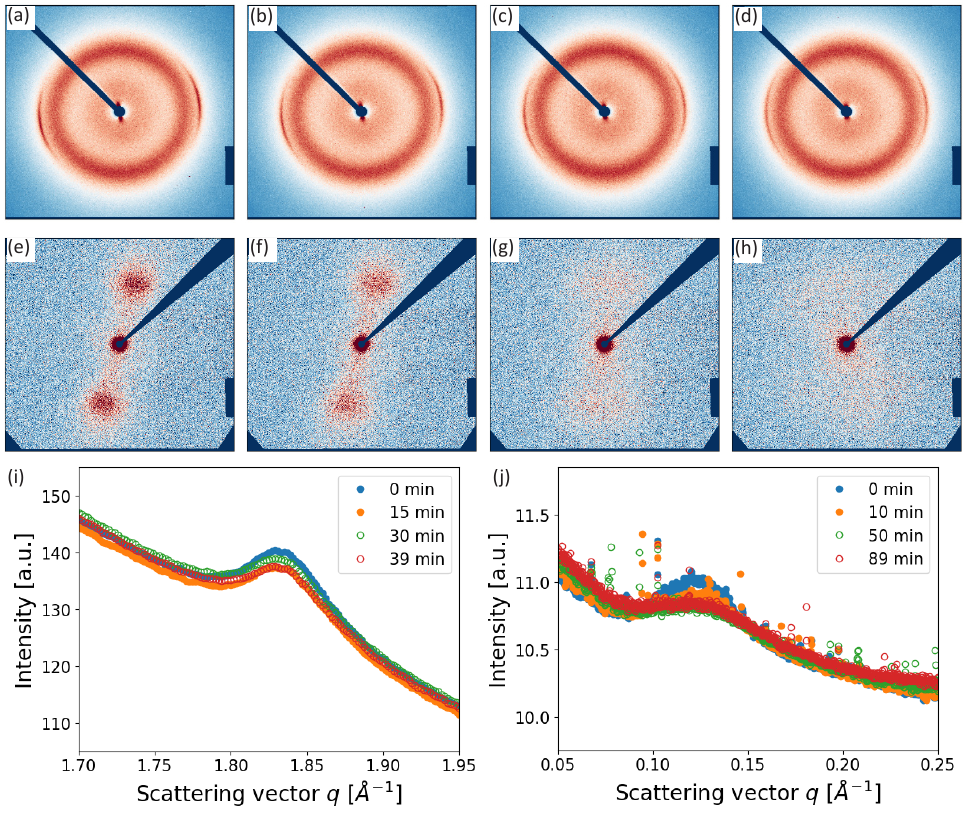}  
  \caption{\label{fig:raw-local} 
Comparison of X-ray scattering data for samples subjected to local UV irradiation. The sample was illuminated only over a localized region. (a–d) In the wide-angle region, the scattering peak broadens while the intensity decreases relative to the initial sample. (e–h) A similar transformation is observed in the small-angle region. Note that the irradiation times used in the WAXD and SAXS measurements are different to achieve comparable UV-induced changes. Accordingly, panels (a–d) correspond to 0, 15, 30, and 39 min, whereas panels (e–h) correspond to 0, 15, 50, and 89 min. (i,j) show the azimuthally averaged intensity profiles.}
\end{figure}
